\documentclass[sigconf]{acmart}
\usepackage[utf8]{inputenc}
\usepackage{subfigure}
\usepackage{amsmath}
\usepackage{amsfonts}
\usepackage{natbib}
\usepackage{graphicx}
\usepackage{multirow}

\AtBeginDocument{%
  \providecommand\BibTeX{{%
    \normalfont B\kern-0.5em{\scshape i\kern-0.25em b}\kern-0.8em\TeX}}}

\setcopyright{acmcopyright}
\copyrightyear{2020}
\acmYear{2020}
\acmDOI{10.1145/3383313.3412222}

\acmConference[RecSys '20]{Fourteenth ACM Conference on Recommender Systems}{September 21--26, 2020}{Virtual Event, Brazil}
\acmBooktitle{Fourteenth ACM Conference on Recommender Systems (RecSys '20), September 21--26, 2020, Virtual Event, Brazil}
\acmPrice{15.00}
\acmISBN{978-1-4503-7583-2/20/09}

\begin{document}

\title{Long-tail Session-based Recommendation}


\author{Siyi Liu}
\affiliation{%
   \institution{University of Electronic Science and Technology of China}
   \city{Cheng Du}
   \state{Si Chuan}
   \country{China}}
\email{ssuiliu.1022@gmail.com}

\author{Yujia Zheng}
\authornote{Corresponding author. Email: yjzheng19@gmail.com}
\affiliation{%
   \institution{University of Electronic Science and Technology of China}
   \city{Cheng Du}
   \state{Si Chuan}
   \country{China}}
\email{yjzheng19@gmail.com}


\begin{abstract}
  Session-based recommendation focuses on the prediction of user actions based on anonymous sessions and is a necessary method in the lack of user historical data. However, none of the existing session-based recommendation methods explicitly takes the long-tail recommendation into consideration, which plays an important role in improving the diversity of recommendation and producing the serendipity. As the distribution of items with long-tail is prevalent in session-based recommendation scenarios (e.g., e-commerce, music, and TV program recommendations), more attention should be put on the long-tail session-based recommendation. In this paper, we propose a novel network architecture, namely TailNet, to improve long-tail recommendation performance, while maintaining competitive accuracy performance compared with other methods. We start by classifying items into short-head (popular) and long-tail (niche) items based on click frequency. Then a novel \emph{preference mechanism} is proposed and applied in TailNet to determine user preference between two types of items, so as to softly adjust and personalize recommendations. Extensive experiments on two real-world datasets verify the superiority of our method compared with state-of-the-art works.
\end{abstract}

\ccsdesc[500]{Information systems~Recommender systems}
\keywords{Session-based recommendation, Long-tail recommendation, Neural network}

\maketitle

\section{Introduction}

Session-based Recommendation System (SRS) has attracted much attention because of numerous application areas in online services (e.g.,  e-commerce, music, and web page navigation). Such recommendation only relies on large, time-ordered action logs of anonymous users to predict the user's next action.  \cite{ludewig2018evaluation,quadrana2018sequence, zheng2019balancing, yu2020tagnn, mi2020memory, mi2020ader}.

Recent studies mainly focus on achieving state-of-the-art accuracy performance, without taking long-tail recommendation into consideration in SRS. 
However, previous studies provide a comprehensive explanation of the significance of long-tail based recommendation from two angles. For users, only recommending popular (short-head) items get them bored easily. Long-tail recommendation can increase the diversity and serendipity of a recommendation list, which surprises and satisfies them because long-tail items can still be very relevant to users \cite{adomavicius2012improving,li2017two}. For business, long-tail items embrace relatively large marginal profit compared with short-head items, which means that long-tail recommendation can bring much more profit \cite{anderson2006long,yin2012challenging}. Besides, long-tail recommendation can give users “one-stop shopping convenience”, which can entice customers to consume both short-head items and long-tail items at one-step, and thereby creating more sales \cite{yin2012challenging,goel2010anatomy}.
    

\begin{figure*}[h]
    \centering
    \includegraphics[width=0.9\textwidth]{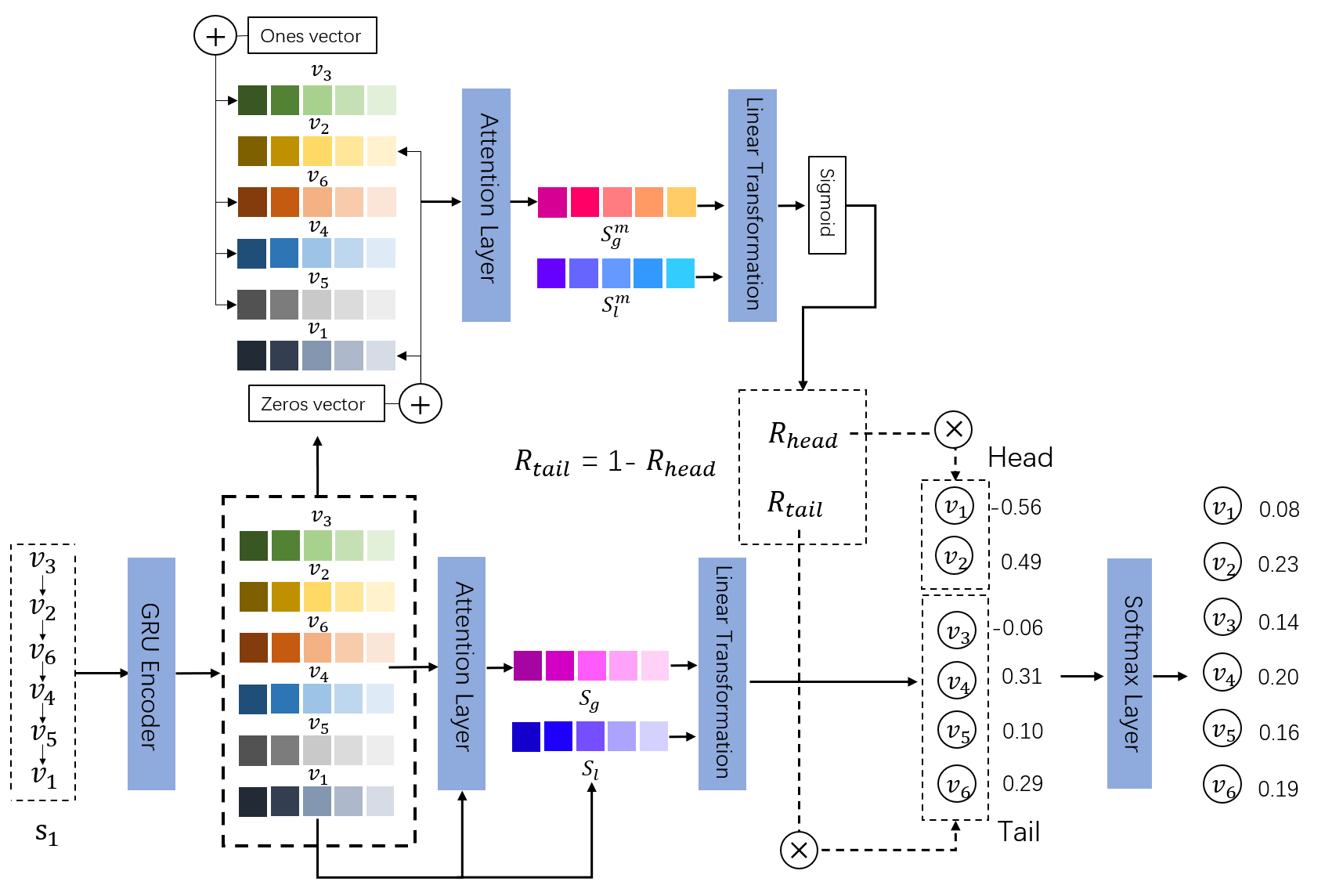}
    \caption{The framework of TailNet. 
    We first encode the session by a GRU layer and get the session's latent representation $V$. Then it is sent to the \emph{preference mechanism} and the attention layer. \emph{Preference mechanism} uses $V$ to generate \emph{rectification factors} $R_{head}$ and $R_{tail}$, which softly adjust the mode of TailNet. Next, the global session embedding $S_g$ is generated by the attention layer. After that, the global embedding $S_g$ and local embedding $S_l$ (the last clicked item's representation) are concatenated and sent to a linear transformation. Finally, we predict the probability of each item with the adjusting of $R_{head}$ and $R_{tail}$. 
    }
    \label{fig2}
\end{figure*}
Obstacles such as data sparsity stand in the way of applying long-tail recommendation, resulting in most existing recommendation systems, including SRS, being biased towards popular items \cite{yin2012challenging,li2017two,shi2013trading}. In these methods, the preference to recommend the most popular items is a conservative but effective way to improve accuracy \cite{adomavicius2012improving}. Previous works focusing on long-tail recommendations either sacrifice the accuracy of recommendations \cite{adomavicius2012improving,shi2013trading,johnson2017enhancing}, or adopt side-information(e.g.,  user profiles, action types), which exceeds the data limit of SRS, to mitigate long-tail items data sparsity \cite{li2017two,bai2017dltsr}.

The difficulty of long-tail recommendation is greater for session-based data. \emph{Firstly,} due to the lack of side information and user profiles in sessions, it is hard for traditional long-tail recommendation methods to be applied in the session-based recommendation.
\emph{Secondly,} most of the traditional long-tail recommendation methods have difficulty in taking the sequential order of data into consideration. Thus, the recommendation accuracy of those methods severely drops when it comes to session-based recommendation.
\emph{Thirdly,} numerous session data is particularly sparse because each user could have several sessions and each session should be handled independently \cite{hidasi2015session}, which deteriorates the model performance on the long-tail recommendation.
Therefore, the development of a new method that can solve the above difficulties is of great significance.

In this paper, to address long-tail phenomenon in SRS, we propose a novel network structure–TailNet. Unlike existing neural networks based recommendation algorithms, our method can significantly improve long-tail recommendation performance while maintaining competitive results with state-of-the-art method simultaneously. In particular, we first encode each session into latent representation. Then we introduce a novel \emph{preference mechanism} into our model, which determines user preference between clicking short-head items and long-tail items by learning from session representation and generate rectification factors. TailNet can utilize the rectification factors given by \emph{preference mechanism} to softly adjust the recommendation mode of the model: recommend more short-head items or long-tail items. The \emph{preference mechanism} is jointly learned with TailNet in an end-to-end back-propagation training paradigm. The main contributions of our work are two fold. First, We propose a novel neural network based model named TailNet which takes into consideration of long-tail phenomenon. Second, we conduct extensive experiments on real-world datasets. Experimental results show that our method achieves state-of-the-art performance on long-tail recommendation as well as relatively high accuracy.

\section{Model}
\subsection{Notations}
The aim of SRS can be defined as using the user's current session data to predict the user's next click item. $I = \{i_1, i_2, ... ,i_{|I|}\}$ represents the set of unique items in all sessions. In this paper, we split items into  $I^H$ and $I^T$ according to the \textbf{Pareto Principle}, which represents the set of short-head items and long-tail items respectively. $s = \{i_1, i_2, ... i_t\}$ represents the session which consists of items ordered by timestamps. $V = \{v_1, v_2, ... ,v_t\}$ represents the latent representation for the session encoded by session encoder layer, where $v_t \in V$ represents a clicked item representation within the $V$ at timestamp $t$. In TailNet, we first get vector $\widehat{c}$, where each element value is the score of the item before softmax. Then we use \emph{preference mechanism} to get \emph{rectification factors} marked as $R_{head}$ and $R_{tail}$ (corresponding to short-head items and long-tail items respectively) for soft adjustment. Finally, the TailNet outputs probabilities $\widehat{y}$ for possible items whereby each element value of vector $\widehat{y}$ is the recommendation score of the corresponding item. 

embedding $S_l^m$ (the last clicked item's representation), then compresses it into a representation $S_p$. After sending $S_p$ to sigmoid function, we can get $R_{head}$ and $R_{tail}$.
We use Gated Recurrent Unit (GRU) \cite{cho2014learning} to encode session $s$, where the GRU is defined as:
\begin{equation} \label{eq1}
  \begin{split}
  r_t &= \sigma(W_r \cdot [ v_{t-1 },emb(i_t)]),   \\
   z_t &= \sigma(W_z \cdot [v_{t-1},emb(i_t)]), \\
  \widetilde{v}_t &= tanh(W_{h}\cdot[r_t \odot v_{t-1},emb(i_t) ]), \\
   v_t &= (1-z_t)\odot v_{t-1} + z_t \odot \widetilde{v}_t,
  \end{split}
\end{equation}
where $W_r$, $W_z$ and $W_h$ denote the weights of the corresponding gates; $emb(i_t)$ denotes the embedding of item $i_t$; $\sigma$ denotes the sigmoid function. We initialize $v_0 = 0$. And each session $s$ is encoded into $V = \{v_1, v_2,  . . . , v_t\}$.
\begin{table*}[h]
\centering
    \scriptsize
    \centering
    \caption{The performance of TailNet with other baseline methods on two datasets. (PM: Preference Mechanism)}\label{tab2}
    \resizebox{\textwidth}{!}{
    \begin{tabular}{c ccccc ccccc}
        \toprule

\multirow{2}{*}{\textbf{Methods}} & \multicolumn{5}{c}{\textbf{30MUSIC}} & \multicolumn{5}{c}{\textbf{YOOCHOOSE 1/4}} \\
\cmidrule(r){2-6} \cmidrule(r){7-11} 

&  MRR@20 &  Recall@20  &  Coverage@20 &  Tail\_Coverage@20  &  Tail@20
&  MRR@20  &  Recall@20 &  Coverage@20  &  Tail\_Coverage@20  &  Tail@20  \\

\midrule
POP  & 0.18 & 0.69 & 0.01 & 0 & 0  
         & 0.3 & 1.36 & 0.06 & 0 & 0 \\
S-POP  & 8.20 & 18.98 & 15.52 & 10.55 & 20.16  
         & 17.85 & 27.18 & 19.59 & 9.29 & 1.82\\
Item-KNN  & 15.71 & 37.68 & 75.14 & 79.04 & 54.60  
         & 21.68 & 53.01 & 65.93 & 60.72 & $14.65^{\dagger}$ \\
FPMC  & 9.17 & 14.47 & $84.37^\dagger$ & $97.45^\dagger$ & $60.00^\dagger$ 
         & 19.17 & 46.69 & $70.97^\dagger$ & $84.74^\dagger$ & 6.83 \\         
BPR-MF  & 6.97 & 12.25 & 49.62 & 86.95 & 19.94 
         & 16.37 & 23.77 & 44.82 & 68.32 & 2.96 \\
\midrule         
GRU4REC & 20.45 & $\textbf{39.12}^\dagger$ & 36.60 & 26.18 & 18.64   
         & 22.41 & 59.58 & 25.79 & 11.71 & 1.87 \\
NARM  & 20.19 & 36.68 & 45.51 & 36.99 & 30.84
         & 28.88 & 69.29 & 44.05 & 30.65 & 6.49 \\
STAMP  & 13.12 & 23.34 & 14.10 & 3.81 & 12.95  
         & 30.33 & 70.55 & 41.11 & 26.89 & 6.34 \\
RepeatNet &18.01 &33.01 &32.24  &24.49  &23.06  
              &$\textbf{31.01}^\dagger$ &70.69 &33.82  &19.00  &4.62  \\
SR-GNN  & 27.28 & 37.86 & 40.59 & 28.71 & 20.48  
         & 30.64 & $\textbf{71.39}^\dagger$ & 33.90 & 19.71 & 6.43 \\

\midrule
TailNet without PM  & 28.62 & 39.00 & 34.53 & 21.33 & 11.56  
            & 30.64 & 69.29 & 42.75 & 28.70 & 6.92 \\
 
TailNet-propotion  & 27.91 & 37.29
        & 38.55 & 25.37 & 19.85  
        & 28.99 & 67.04 & 43.86 & 30.32 & 8.43 \\           

TailNet  & $\textbf{28.70}^\dagger$ & 38.34
            & \textbf{47.86} & \textbf{40.10} & \textbf{32.13}  
            & 30.97 & 69.41 & \textbf{45.51} & \textbf{32.31} & \textbf{8.88} \\
\bottomrule
\end{tabular}}
\footnotesize
    \textbf{Boldface} indicates the best result in neural network based methods.
    \textbf{$\dagger$} indicates the best results in all methods. The scores reported in \cite{wu2018session} on the YOOCHOOSE dataset differ because of the discrepancy in test data. For neural network based methods, we choose the values of Coverage, Tail\_Coverage and Tail when they are most accurate.

\end{table*}

\begin{figure*}[h]
\centering
\includegraphics[width=0.9\textwidth]{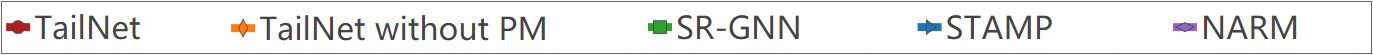}
\subfigure[\textbf{Coverage@K}]{
\begin{minipage}[b]{0.3\textwidth}
\includegraphics[width=1\textwidth]{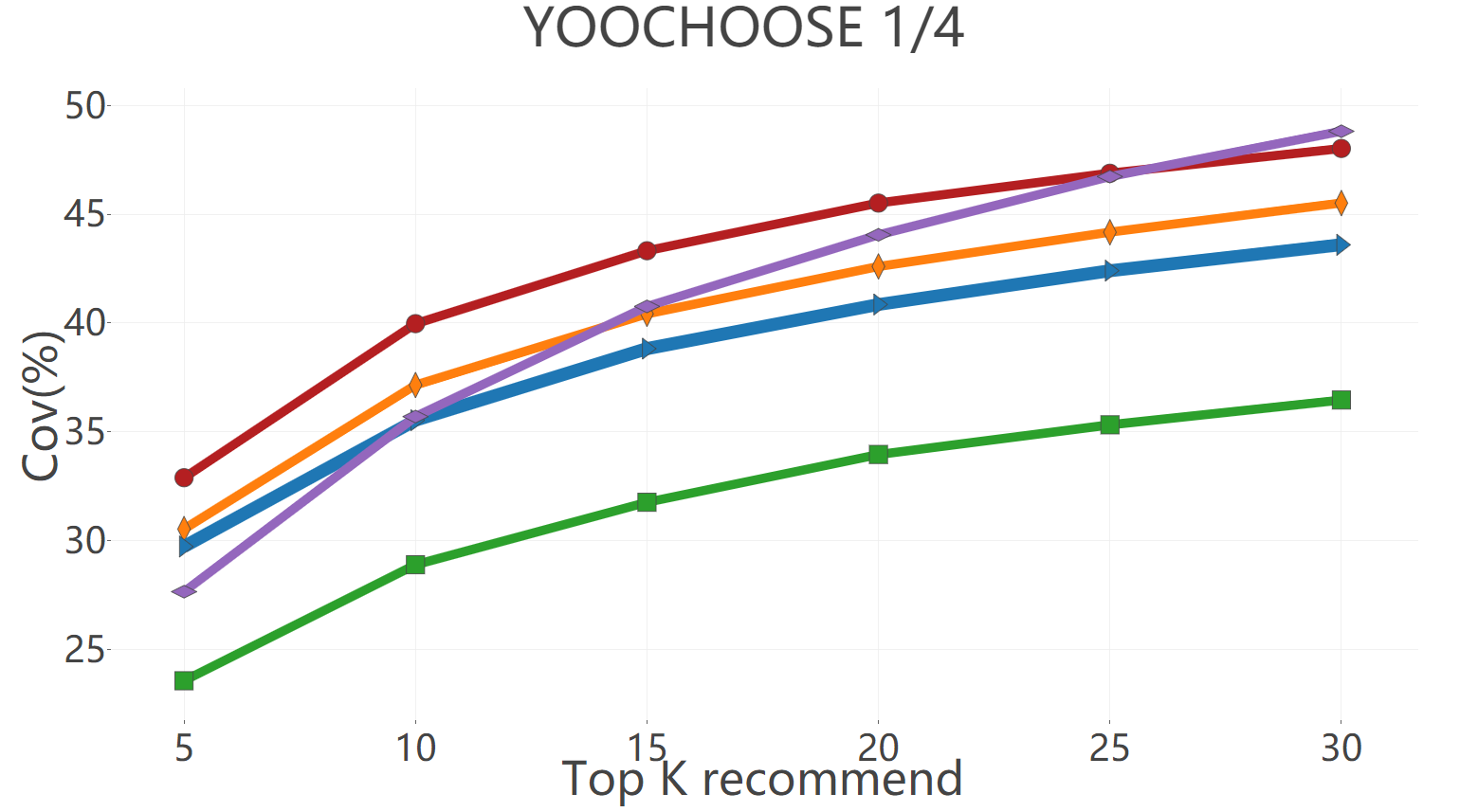} \\
\includegraphics[width=1\textwidth]{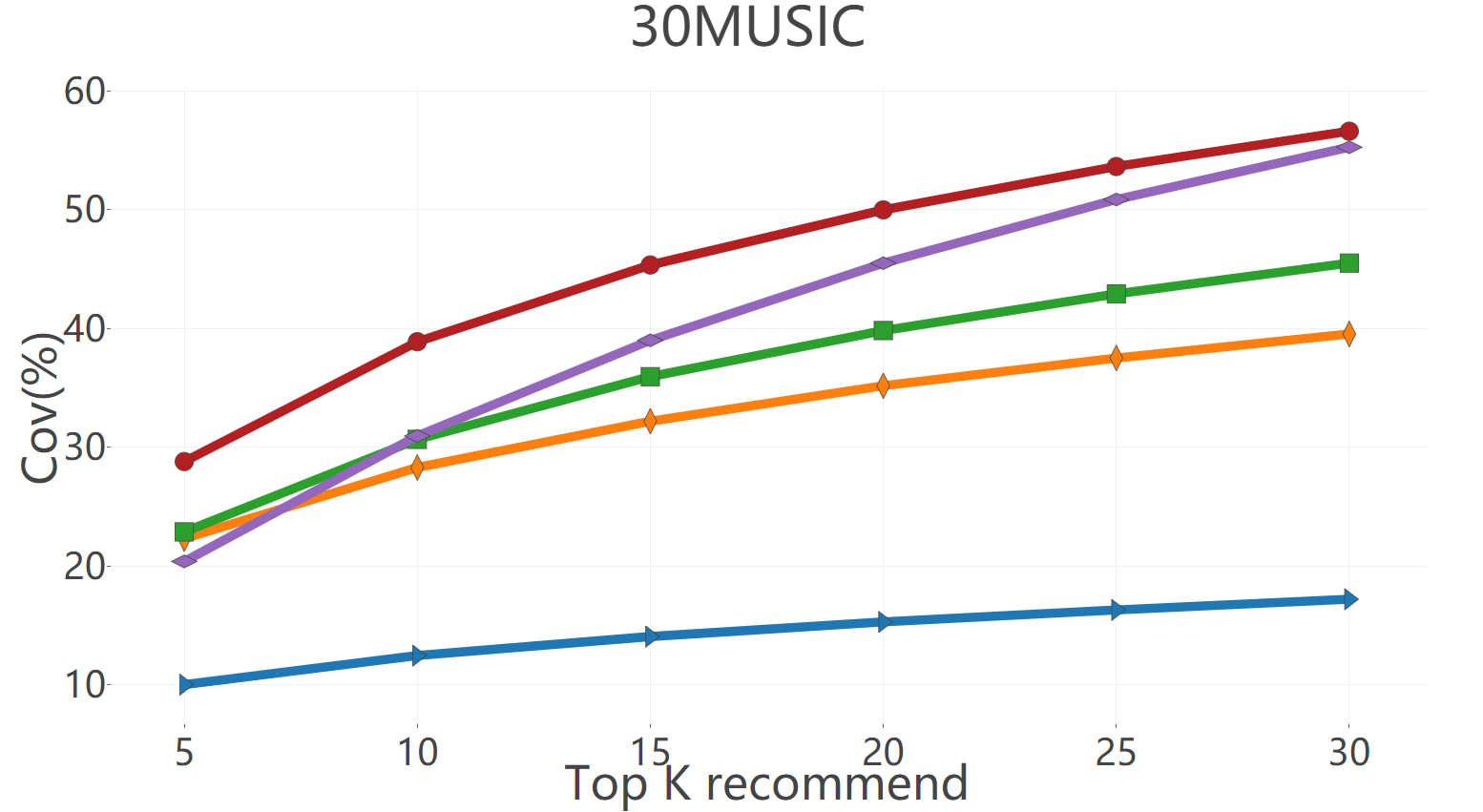}
\end{minipage}
}
\subfigure[\textbf{Tail@K}]{
\begin{minipage}[b]{0.3\textwidth}
\includegraphics[width=1\textwidth]{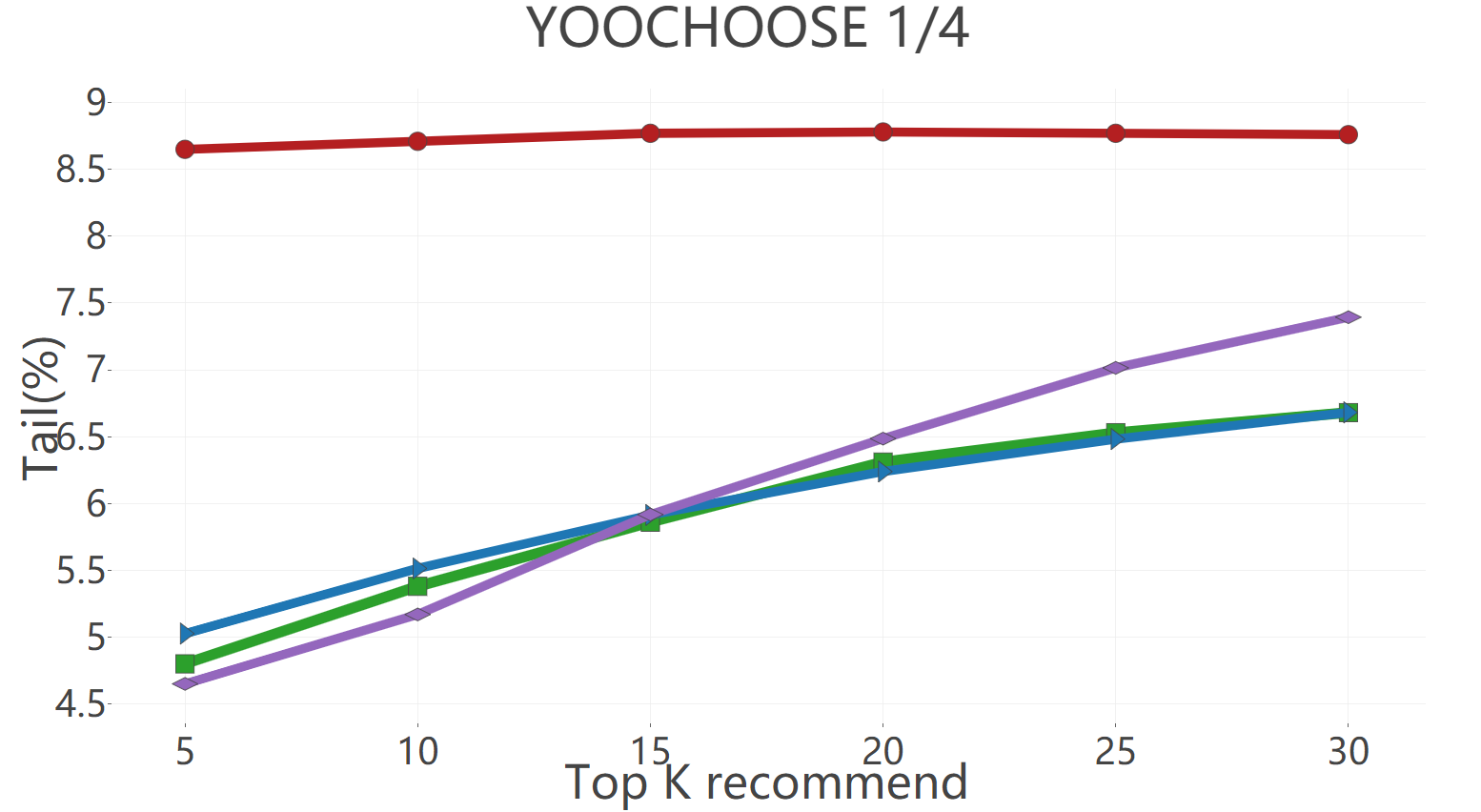} \\
\includegraphics[width=1\textwidth]{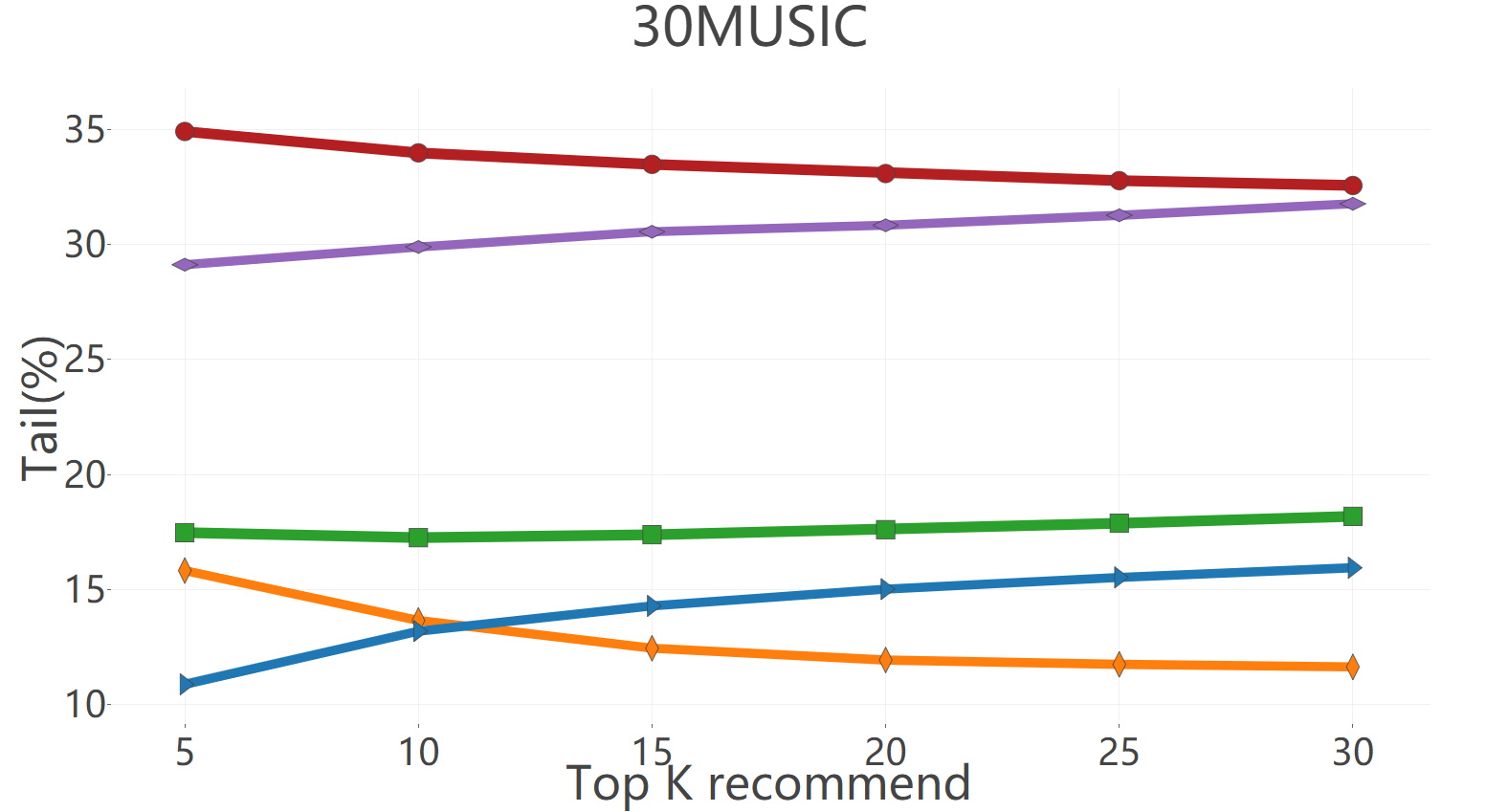}
\end{minipage}
}
\subfigure[\textbf{Tail\_Coverage@K}]{
\begin{minipage}[b]{0.3\textwidth}
\includegraphics[width=1\textwidth]{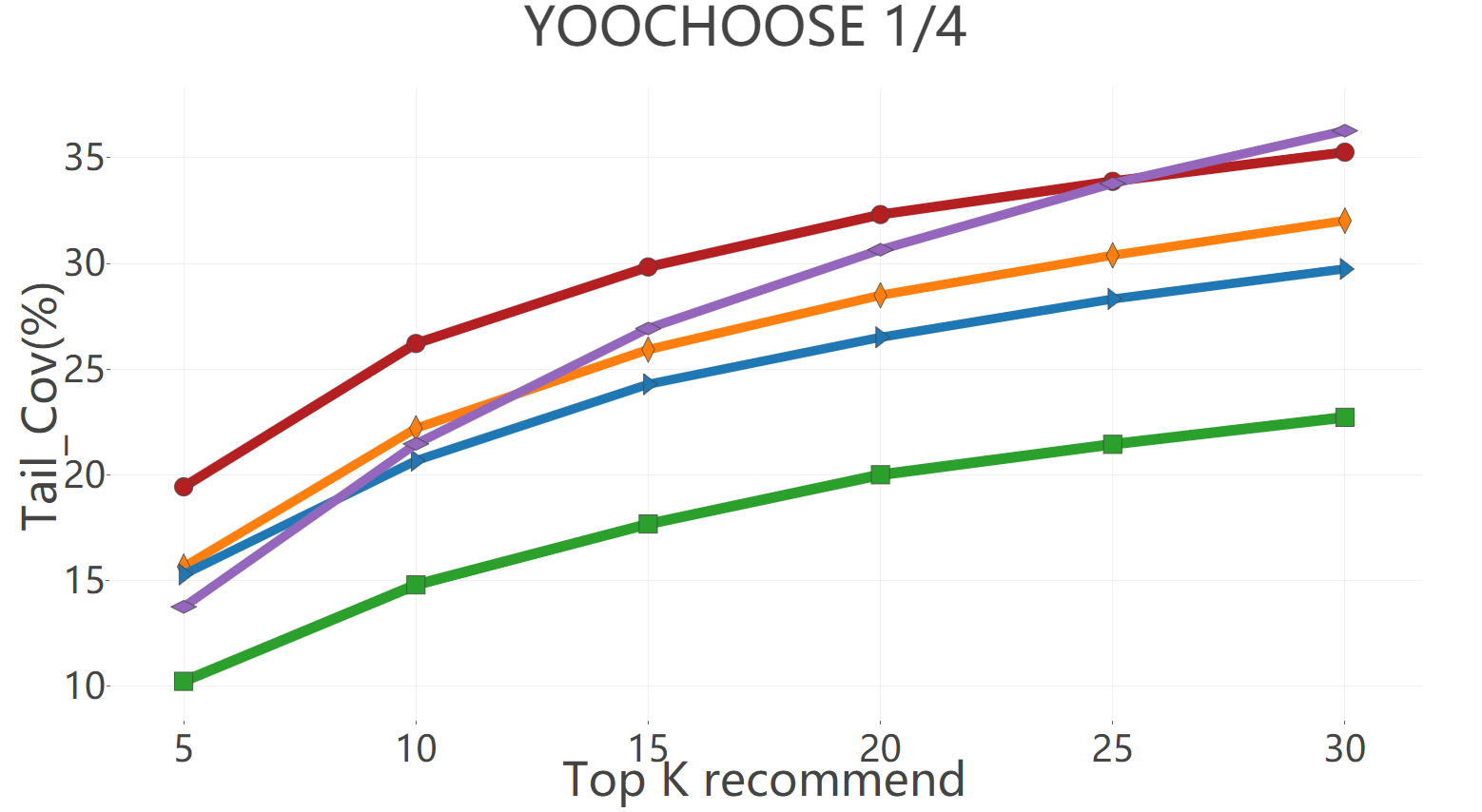} \\
\includegraphics[width=1\textwidth]{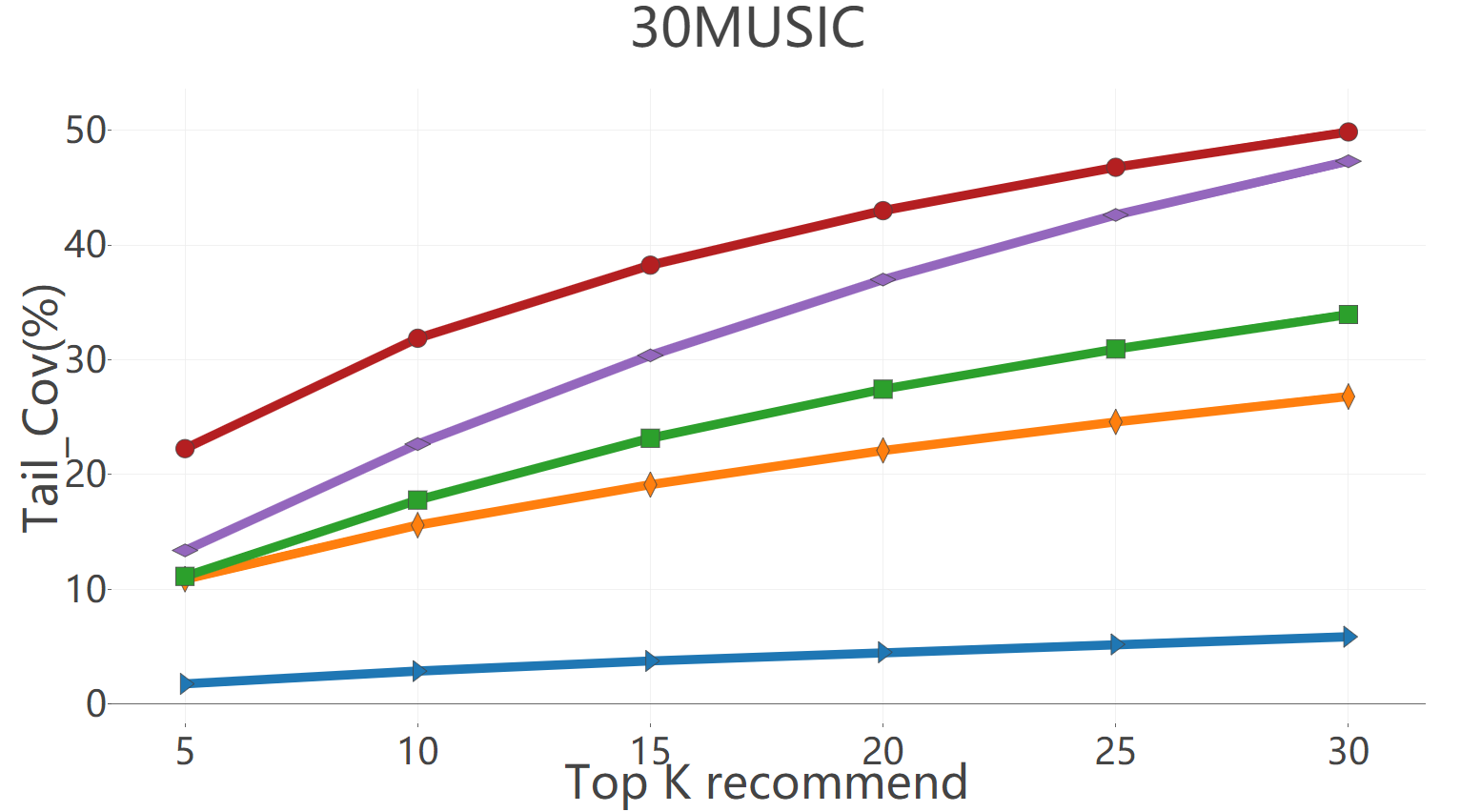}
\end{minipage}
}

\caption{Long-tail recommendation performance comparison between TailNet and state of the art baseline methods on two datasets. (PM: Preference Mechanism)}
\label{fig4}
\end{figure*}
\subsection{Preference Mechanism}
We introduce a \emph{preference mechanism} to softly adjust the mode of TailNet. As illustrated in Figure \ref{fig2}, The \emph{preference mechanism} contains an encoder, a linear transformation layer with attention mechanism and a sigmoid function. The mechanism takes session's latent representations $V = \{v_1, v_2, ... ,v_t\}$ as an input. 

First, we need to embed the item type into the item representation. For inputs of long-tail and short-head items (splitted by \textbf{Pareto Principle} \cite{anderson2006long}), we add a vector of ones and a vector of zeros, respectively. However, making these vectors adjustable should distinguish the popularity bias more flexibility. To most directly illustrate the effectiveness of the proposed mechanism, we use vectors of ones and zeros here:
\begin{equation} \label{eq2}
  \begin{split}
  TE(v_i)=\left\{
\begin{array}{rcl}
v_i + \{1,1,... ,1\}       &      & {{\rm if}\ \tau(v_i) \in I^T}\\
v_i + \{0,0,... ,0\}       &      & {{\rm if}\ \tau(v_i) \in I^H},
\end{array} \right. 
  \end{split}
\end{equation}
where $\tau(v_i)$ denotes a mapping function, $\tau(v_i)$ maps $v_i$ to corresponding item $i$. The $ones$ and $zeros$ vector have the same size as the input, i.e.$ones$, $zeros \in \mathbb{R}^{d}$. $d$ is the dimension of the inputs.  
\label{sec3.4.1}

Then we adopt the attention mechanism to represent the session preference $S_p$:
\begin{equation} \label{eq3}
  \begin{split}
  \alpha^m_i &= W^m_0tanh(W^m_1TE(v_t)+W^m_2TE(v_i) + b^m), \\
  S^m_l &= TE(v_t), \\
  S^m_g &= \sum_{i=1}^{t}\alpha^m_iTE(v_i),    \\
  S_p &= [S^m_l;S^m_g]W_3,
  \end{split}
\end{equation}
where $ TE(v_i), TE(v_t) \in \mathbb{R}^d $ denote the representation of item $i$ and the last-cliked item after Tail Encoder respectively; $ W^m_0 \in \mathbb{R}^{1 \times d}$ and $ W^m_1,W^m_2 \in \mathbb{R}^{d \times d}$ are weights in the attention mechanism of \emph{preference mechanism}; matrix $ W_3 \in \mathbb{R}^{2d \times 1}$ compresses two combined embedding vectors into the latent space $ \mathbb{R}^1$; $b^m \in  \mathbb{R}^d $ is a bias vector; $\alpha^m _i$ represents each item $i$'s attention weight coefficient. 

%
After that, a sigmoid function is ued to $S_p$ to get \emph{rectification factors}:
\begin{equation} \label{eq5}
  \begin{split}
     R_{head} &= \frac{1}{1+e^{-S_p}}, \\
     R_{tail} &= 1- R_{head},
  \end{split}
\end{equation}
where $R_{head}$ and $R_{tail}$ denote the factors corresponding to short-head items and long-tail items respectively.

\subsection{Session Pooling and Soft Adjustment}

Same as the \emph{preference mechanism} above, an attention layer and a linear transformation are used to process latent representations $V$:
\begin{equation} \label{eq6}
  \begin{split}
  \alpha_i &= W_0tanh(W_1v_t+W_2v_i + b), \\
  S_l &= v_t, \\
  S_g &= \sum_{i=1}^{t}\alpha_iv_i,    \\
  \widehat{c} &= [S_l;S_g]W_4,
  \end{split}
\end{equation}
%
%
 where $W_0$, $W_1$, $W_2$, $W_4$ are weight matrices and $b$ is the bias vector. $S_g$ denotes global embedding and $v_t$ denotes local embedding. Different from $S^m_g$ and $S^m_l$\ in \emph{preference mechanism}, $S_g$ and $S_l$ represent user's general and current interests in next-click item. Each element's value of  $\widehat{c}=\{ \widehat{c}_1,\widehat{c}_2,...,\widehat{c}_{|I|}\}$ is the score of the corresponding item before softmax.

Finally, to adjust recommendation mode, we multiply each $ \widehat{c}_i$ with corresponding \emph{rectification factor} $(R_{head}\ or\ R_{tail})$, then apply softmax function on it to get probabilities for recommendation:
\begin{equation} \label{eq8}
  \begin{split}
 \widehat{y} &= softmax(\widehat{c} \odot R ),
  \end{split}
\end{equation}
where $R \in \mathbb{R}^{|I|} $  is a vector, and each element's value of it is the $R_{head}$ or $R_{tail}$ corresponding to the item in head or in tail, e.g., $R=\{R_{head},R_{tail},...,R_{tail}\}$. $\odot$ denotes the element wise product. Each element's value of vector $\widehat{y}\in \mathbb{R}^{|I|}$ is the recommendation score of the corresponding item. The items with top-K values in vector value will be recommended as the final output. In addition, TailNet is trained jointly with \emph{preference mechanism} by Back-Propagation Through Time (BPTT) \cite{werbos1990backpropagation}. We use the sum of the cross-entropy of each item's prediction and the ground truth as the loss: 
\begin{equation} \label{eq9}
  \begin{split}
  \mathcal{L}(\hat{\mathbf{y}})=-\sum_{i=1}^{|I|} \mathbf{y}_{i} \log \left(\hat{\mathbf{y}}_{i}\right)+\left(1-\mathbf{y}_{i}\right) \log \left(1-\hat{\mathbf{y}}_{i}\right),
  \end{split}
\end{equation}
where $\mathbf{y}$ denotes the one-hot encoding vector of the ground truth item.

\section{Experiments}

We evaluate our proposed method on two real-world datasets: YOOCHOOSE\footnote{\url{http://2015.recsyschallenge.com/challenge.html}} and 30MUSIC\cite{turrin201530music}.
For a fair comparison, we use the same data preprocessing approach as previous studies \cite{liu2018stamp, wu2018session}. We use the recent fractions 1/4 of the YOOCHOOSE datasets as \cite{tan2016improved}. For both datasets, we use the last day to generate test data. Because of collaborative filtering methods cannot recommend an item which has not appeared before \cite{hidasi2015session}, we filter out that kind of items in test data. According to \textbf{Pareto Principle} \cite{anderson2006long}, we split the itemset into short-head items and long-tail items.

\textbf{Baseline. }
We compare TailNet with frequency based methods (POP and S-POP), two RNN-based methods (GRU4REC \citep{hidasi2015session} and NARM \citep{li2017neural}), neighborhood-based method (Item-KNN \cite{sarwar2001item}), repeat-explore method (RepeatNet \cite{ren2018repeatnet}), attention-based method (STAMP \cite{liu2018stamp}), two traditional Matrix Factorization or Markov Chain approaches (FPMC \citep{rendle2010factorizing} and BPR-MF \citep{rendle2009bpr}), and GNN-based method (SR-GNN \citep{wu2018session}).

\renewcommand{\thesubtable}{\Roman{subtable}}
\renewcommand{\thesubfigure}{\arabic{subfigure}}

\textbf{Evaluation Metrics. }
We use Recall@K and MRR@K to evaluate the performance of all algorithms in terms of accuracy. And for the evaluation of long-tail recommendation , we employ three diversity metrics into the evaluation.
\begin{itemize}
    \item {\textbf{Coverage@K and Tail-Coverage@K}: 
    Coverage@K and Tail-Coverage@K measure how many different items and long-tail items ever appear in the top-K recommendations respectively. The Coverage@K and Tail-Coverage@K are defined as:

    \begin{equation} \label{eq10}
        \begin{split}
            Coverage@K &= \frac{|\cup_{s\subset S}L_K(s)|}{|I|}, \\
            Tail\_Coverage@K &= \frac{|\cup_{s\subset S}L_K^T(s)|}{|I^T|},
        \end{split}
    \end{equation}
    where $L_K(s) = [i_1, i_2, ... ,i_k]$ represents the list of top-K recommended items for session $s$ and $L_K^T(s)$ represents a subset of $L_K(s)$ which contains items that belong to $|I^T|$
    }
    \item {\textbf{Tail@K}: 
    Tail@K measure how many long-tail items in top-K for each recommendation list. The overall Tail@K is defined by averaging all test cases:
    \begin{equation} \label{eq12}
        \begin{split}
            Tail@K = \frac{1}{|S|}\sum_{s\in S} \frac{|L_K^T(s)|}{K}.
        \end{split}
    \end{equation}
    }
\end{itemize}

\section{Results and Analysis}

\subsection{Comparison with Baseline Methods}

The results of TailNet and all baseline methods over two datasets are presented in Table \ref{tab2}, and a more specific comparison between TailNet and other neural network based methods in top-K recommendation is shown in Figure \ref{fig4}. The accuracy of traditional methods is relatively lower than that of neural network based methods. However, most of the traditional methods achieve better performance on long-tail recommendation than neural network based methods. This result indicates that only using accuracy to evaluate the quality of a recommendation algorithm may not be sufficient. Moreover, the trade-off between accuracy and long-tail recommendation highlights the challenge of balancing both metrics. In both datasets, TailNet achieves the best long-tail recommendation performance among state-of-the-art deep learning methods and the best accuracy among traditional methods, which confirms that our proposed method makes more comprehensive recommendations. Obviously, none of the baseline methods can balance long-tail recommendations and accurate recommendations as well as TailNet. Although our work mainly focuses on tackling long-tail recommendation, TailNet still achieves comparable results in terms of MRR@20, Recall@20. This result demonstrates that the application of \emph{preference mechanism} only makes a bit of fluctuation in accuracy. \emph{Preference mechanism} can accurately determine user preference between long-tail items and short-head items and adjust the recommendation result properly, not just rigidly recommend more niche items regardless of user preference.

\subsection{The Effect of Soft Adjustment}
TailNet can softly adjust the recommendation scores with the \emph{rectification factors} generated by \emph{preference mechanism}. However, because we divide short-head items and long-tail items according to \textbf{Pareto Principle}, we cannot exclude the possibility that the competitive performance of TailNet is largely due to this prior knowledge. Thus, we design a hard adjustment version of TailNet, namely TailNet-proportion, to study the effect of model based only on \textbf{Pareto Principle}. Specifically, for top-$k$ recommendations, TailNet-proportion generates the recommendation list by combining the top $[k*p]$ short-head items and the top $k-[k*p]$ long-tail items, then sort them
based on the scores. $p$ represents the proportion of short-head items in a session and $[\cdot]$ indicates the rounding function. From the Table \ref{tab2} we can observe that TailNet outperforms TailNet-proportion in all metrics, which verifies that soft adjustment rather than \textbf{Pareto principle} plays a major role in TailNet.








\subsection{Application of Preference Mechanism in Other Models}

Our proposed \emph{preference mechanism} can be easily integrated with other neural network based model to improve the performance of long-tail recommendation (The mechanism only needs to take session's latent representation generated by the model as an input, and then outputs \emph{rectification factors} $R_{head}$ and $R_{tail}$). To verify that, we integrate the \emph{preference mechanism} into two states of the art baseline models: STAMP and SR-GNN, and make comparisons between those models with and without \emph{preference mechanism}. As shown in Table \ref{tab5}, STAMP and SR-GNN obtain significant improvement after applying the \emph{preference mechanism}, which verifies the portability of the \emph{preference mechanism}. By adopting the \emph{preference mechanism}, neural network based methods can overcome information insufficiency obstacles in long-tail recommendations. Similar to TailNet’s performance, the \emph{preference mechanism} improves the longtail recommendation of STAMP and SR-GNN with a small range of accuracy fluctuation, which further validates the effectiveness of softly adjusting recommendation results according to user preferences.

\begin{table}
\centering
    \scriptsize
    \centering
    \caption{The performance of STAMP and SR-GNN,  both with and without preference mechanism, on YOOCHOOSE 1/4 dataset. (PM: Preference Mechanism)}\label{tab5}
    \resizebox{0.45\textwidth}{!}{
    \begin{tabular}{c ccccc ccccc}
        \toprule

\multirow{2}{*}{\textbf{Methods}} & \multicolumn{5}{c}{\textbf{@20(\%)}} & \\
\cmidrule(r){2-6} 

&  MRR  &  Recall &  Coverage  &  Tail\_Coverage  &  Tail  \\

\midrule
STAMP  & 30.33 & \textbf{70.55} & 41.11 & 26.89 & 6.34 \\
STAMP with PM & \textbf{30.83} & 70.46 & \textbf{43.80} & \textbf{30.33} & \textbf{7.08}\\
\midrule
SR-GNN  & \textbf{30.64} & \textbf{71.39} & 33.90 & 19.71 & 6.43  \\
SR-GNN with PM & 30.62 & 71.37 & \textbf{35.46} & \textbf{21.99} & \textbf{6.57}  \\
\bottomrule

\end{tabular}}

\end{table}


\bibliographystyle{ACM-Reference-Format}
\bibliography{sample-sigconf}

\end{document}